\documentclass{article}
\usepackage{graphics}
\usepackage{setspace}

\begin{document}

\title{\noindent Electron-Phonon Dynamics in an Ensemble of Nearly Isolated Nanoparticles}

\author{Daniel T. Simon and Michael R. Geller \\
\textit{Department of Physics and Astronomy}\\
\textit{University of Georgia, Athens, GA 30602-2451, USA}}
\maketitle

\begin{abstract}

We investigate the electron population dynamics in an ensemble of nearly isolated 
insulating nanoparticles, each nanoparticle modeled as an electronic two-level 
system coupled to a single vibrational mode. We find that at short times the ensemble-averaged 
excited-state population oscillates but has a decaying envelope. At long 
times, the oscillations become purely sinusoidal about a ``plateau'' population,
with a frequency determined by the electron-phonon interaction 
strength, and with an envelope that decays algebraically as $t^{-{1/2}}$. We use this
theory to predict electron-phonon dynamics in an ensemble of Y$_2$O$_3$:Eu$^{3+}$ 
nanoparticles.

\end{abstract}

\section{Introduction}

Nanocrystals exist in a size regime which lies between that of atomic and bulk
matter, thus making them ideal for the study of extreme quantum effects in condensed
matter systems. In particular, the vibrational properties of nanoparticles are
strikingly different than their bulk counterparts: A spherical nanoparticle
of diameter $d$ cannot support internal vibrations at frequencies less
than about $2\pi v/d$, where $v$ is a characteristic bulk sound velocity.
Any property of the nanoparticle that depends on the vibrational spectrum, such
as its thermodynamic properties or electron-phonon dynamics, will be very different
at low energies than in bulk crystals. This will be especially true for 
nanoparticles---for example, in powder form---only weakly coupled to their surroundings.

One way to probe the vibrational spectrum of a nanoparticle is to optically
excite an electron-hole pair and study the intraband electronic energy relaxation
prior to radiative recombination \cite{Woggon,Gaponenko}. However, the excitonic states, 
being only weakly localized, will suffer significant quantum-confinement effects in 
the nanoparticle, making comparison with bulk relaxation rather indirect. An alternative 
probe of the vibrational spectrum is provided by well-localized electronic impurity states
in a doped nanocrystal \cite{Yang etal,Meltzer and Hong}. The impurity states can be used
to probe both energy relaxation by phonon emission \cite{Yang etal} and phonon-induced
dephasing \cite{Meltzer and Hong}. In these cases, the difference between the nanocrystal 
and bulk behavior is almost entirely a consequence of their differing vibrational modes.

In a recent experiment, Yang {\it et al.}~\cite{Yang etal} measured the phonon 
emission rate in Eu-doped Y$_2$O$_3$ nanoparticles between two electronic states 
separated by 3~cm$^{-1}$ in energy. The experiment was performed on a powder of nanoparticles, 
prepared by gas-phase condensation, with a mean diameter of 13~nm. Regarding the
nanoparticles as isotropic elastic spheres \cite{Lamb}, the lowest internal vibrational 
mode (a five-fold degenerate torsional mode) has a frequency of about 11~cm$^{-1}$.
The excited-state population was found to decay exponentially with a lifetime of 27~$\mu$s, 
more than two orders of magnitude longer than that between the same levels in the 
bulk (about 220~ns). Although there is no quantitative theory available yet to explain 
their results, the inhibited phonon emission is consistent with a large suppression of 
the low-frequency vibrational density-of-states (DOS) expected for these small crystals.

Experiments such as the one by Yang \emph{et al.}~\cite{Yang etal} present an exciting
opportunity to study the crossover of phonon dynamics from bulk to nanoscale systems. 
However, the following question immediately arises: Given the large apparent modification 
of the DOS by finite-size effects, is it still correct to expect exponential relaxation
and to use perturbation theory (Fermi's golden rule and so on) to relate the relaxation
rate to the phonon DOS? After all, the energy in a nanoparticle that is completely 
isolated from its surroundings would have to be exchanged between the electron and 
vibrational degrees of freedom in a Rabi-like manner, and no relaxation would be observed. 
Indeed, an isolated nanoparticle may be regarded as a phonon analog of a two-level atom in 
a cavity, which is known to exhibit oscillatory population dynamics \cite{Scully and 
Zubairy}.

It should be possible to detect oscillatory population dynamics experimentally using the
techniques of Ref.~\cite{Yang etal}, but what would be the effect of the unavoidable 
{\it distribution} of nanoparticle sizes and corresponding vibrational-mode frequencies? 
In the limit where each nanoparticle has a two-level system interacting with only the 
lowest-frequency vibrational mode, called the Lamb mode, each nanoparticle would exhibit 
vacuum Rabi oscillations \cite{vacuum footnote}. 
However, the Rabi frequency, which is a function of the 
electron-phonon interaction strength and the energy mismatch between the two-level system 
and Lamb mode, would vary from nanoparticle to nanoparticle.

The purpose of this paper is to investigate the electron-phonon dynamics of an ensemble 
of nearly isolated nanoparticles. We find that at short times the ensemble-averaged
excited-state population oscillates in a fashion that depends on the detailed size
distribution of the nanoparticles, but that at long times the oscillations become purely 
sinusoidal about a ``plateau'' population fraction, with the frequency of the oscillation
determined by the electron-phonon interaction strength alone, and with an envelope that 
decays algebraically as $t^{-{1/2}}$. In the infinite-time limit, the excited-state 
population approaches the constant, finite plateau value. (Of course, the small damping
of the Lamb mode, produced by the weak but nonzero interaction of a nanoparticle with
its environment, will eventually cause the electrons to relax irreversibly.)

In the next section we derive a general expression for the average excited-state 
population in an ensemble of nanoparticles at zero temperature, and study its short-time 
and long-time behavior. The rotating-wave-approximation (RWA) we use there limits the
application of our results to situations where the detuning energies (the energy 
mismatch between the two-level system and Lamb mode) are not much larger than the 
electron-phonon interaction strength. In Section 3 we apply this theory to an ensemble 
of Y$_2$O$_3$:Eu$^{3+}$ nanoparticles similar to that of Yang {\it et al.}~\cite{Yang 
etal}, and give quantitative predictions for the average excited-state population as a
function of the mean nanoparticle size and standard deviation. Section 4 contains our
conclusions and a discussion of questions for future investigation.

\section{Ensemble-averaged population dynamics}

To begin, we consider a single doped nanoparticle. We assume this nanoparticle
to be an isotropic elastic sphere with a single localized electronic
two-level system embedded within. We limit our investigation to small detuning 
energies, thus allowing for the use of the RWA, which discards non-energy-conserving 
terms in the Hamiltonian \cite{Scully and Zubairy}.
In addition, we will neglect the five-fold degeneracy of the lowest vibrational mode
that would be present in a perfectly spherical nanoparticle \cite{Lamb}, and will
assume a single nondegenerate Lamb mode. We shall return to this point below in 
Section 4.

``Dissipation terms'' in the Hamiltonian are neglected as well. By dissipation terms
we mean interactions of the nanoparticle with its surroundings that allow
vibrational energy to be carried away irreversibly. For weak dissipation,
these terms would cause an exponential decay of the population at large times, 
causing the ``plateau'' population to fall exponentially to zero.  This effect would be 
negligible at small times and therefore would have no impact on the short-time 
calculations given below. For large dissipation, the exponential decay would be 
noticeable at all times and would therefore render our results invalid.

We assume that there is only one vibrational mode available for the 
electronic system to couple to---the next vibrational mode being so high in energy
 as to make the effect of its coupling negligible. Therefore, as stated, our doped 
nanoparticle system consists of a two-level atom coupled to single vibrational mode. 
The Hamiltonian, in units where $\hbar=1$, is given by

\begin{equation}
H=\sum _{\alpha }\epsilon _{\alpha }c_{\alpha }^{\dagger }c_{\alpha }+\omega _{0}a^{\dagger }a+\sum _{\alpha \alpha \prime }g_{\alpha \alpha \prime }c_{\alpha }^{\dagger }c_{\alpha \prime }(a+a^{\dagger }),
\label{hamiltonian}
\end{equation}

\noindent where $\alpha=1,2$. The first term in $H$ is the Hamiltonian
for a noninteracting two-level system with energies $\epsilon_{1}$ and
$\epsilon_{2}$, and fermionic creation and annihilation operators $c_{\alpha}^{\dagger}$
and $c_{\alpha}$. The second term is the Hamiltonian for a vibrational
mode with frequency $\omega_{0}$ and phonon creation and annihilation
operators $a^{\dagger}$ and $a$. The third term is the ordinary first-order
interaction between the two-level system and the vibrational mode. $g_{\alpha\alpha\prime}$
is the electron-phonon interaction energy, with $g_{12}=g_{21}=g$, and
the other terms equal to zero.

The wave function can be written as a superposition of the two electronic states,
together with all possible populations of the single vibrational mode,

\begin{equation}
\left| \psi (t)\right\rangle =\sum _{\alpha n}C_{\alpha n}(t)e^{-i(\epsilon _{\alpha }+n\omega _{0})t}\left| \alpha n\right\rangle ,
\end{equation}

\noindent where 

\noindent \begin{equation}
\left| \alpha n\right\rangle \equiv \frac{1}{\sqrt{n!}}(a^{\dagger })^{n}c_{\alpha }^{\dagger }\left| 0\right\rangle .
\end{equation}

\noindent The coefficients $C_{\alpha n}(t)$ satisfy the coupled equations,

\begin{equation}
\partial _{t}C_{1n}(t)+ig\sqrt{n+1}\, e^{-i(\omega _{0}+\Delta \epsilon )t}C_{2,n+1}(t)+ig\sqrt{n}e^{i(\omega _{0}-\Delta \epsilon )t}C_{2,n-1}(t)=0,
\end{equation}

\begin{equation}
\partial _{t}C_{2n}(t)+ig\sqrt{n+1}\, e^{-i(\omega _{0}-\Delta \epsilon )t}C_{1,n+1}(t)+ig\sqrt{n}e^{i(\omega _{0}+\Delta \epsilon )t}C_{1,n-1}(t)=0,
\end{equation}

\noindent where $\Delta\epsilon \equiv \epsilon_{2}-\epsilon_{1}$ is
the electronic energy-level separation. 

In the RWA, which is valid near resonance (defined by the condition
$\omega_{0}=\Delta\epsilon$), these coupled differential equations reduce to

\begin{equation}
\partial _{t}C_{1n}(t)+ig\sqrt{n}\, e^{i\nu t}C_{2,n-1}(t)=0,
\end{equation}

\noindent and

\begin{equation}
\partial _{t}C_{2n}(t)+ig\sqrt{n+1}\, e^{-i\nu t}C_{1,n+1}(t)=0,
\end{equation}

\noindent where 

\noindent \begin{equation}
\nu \equiv \omega _{0}-\Delta \epsilon 
\end{equation}

\noindent is the detuning frequency.

These equations can be solved by Laplace transformation using the boundary condition
$C_{\alpha n}(0)=\delta_{\alpha 2}\delta_{n0}$. The amplitude for the upper state is

\begin{equation}
C_{2n}(t)=\delta _{n0}\left[ \cos \left( \frac{\Omega }{2}t\right) +i\frac{\nu }{\Omega }\sin \left( \frac{\Omega }{2}t\right) \right] e^{-i\nu t/2},
\end{equation}

\noindent where

\begin{equation}
\Omega \equiv \sqrt{\nu ^{2}+4g^{2}}
\end{equation}

\noindent plays the role of the Rabi frequency in this problem \cite{Scully and Zubairy}.

The probability that the electron is in state $\alpha$, irrespective of
the number of phonons present, is

\begin{equation}
N_{\alpha }(t)\equiv \sum _{n}\left| C_{\alpha n}(t)\right| ^{2}.
\end{equation}

\noindent Then \( N_{2}(t)=\left| C_{20}(t)\right| ^{2} \) is given by

\begin{equation}
N_{2}(t)=1-\left( \frac{\Omega ^{2}-\nu ^{2}}{\Omega ^{2}}\right) \sin ^{2}\left( \frac{\Omega }{2}t\right) ,
\end{equation}

\noindent and

\begin{equation}
N_{1}(t)=1-N_{2}(t).
\end{equation}

\noindent The dependence of $N_{1}(t)$ and $N_{2}(t)$ on $\nu$
will be suppressed for simplicity.

In an ensemble of such nanoparticles, a variation in diameter yields a similar
variation in detuning frequency, $\nu$. We assume this to be a Gaussian distribution.
To obtain values for the mean detuning $\overline{\nu }$,
and standard deviation in detuning $\sigma$, we assume each nanoparticle
to be an isotropic elastic continuum with stress-free boundary conditions.
Then, as shown by Lamb \cite{Lamb}, the lowest vibrational frequency is

\begin{equation}
\omega _{0}(d)\approx \frac{2\pi v_{\rm t}}{d},
\label{Lamb mode}
\end{equation}

\noindent where $v_{\rm t}$ is the bulk transverse sound velocity
and $d$ is the nanoparticle diameter. The mean detuning is then

\begin{equation}
\overline{\nu }=\omega _{0}\! \left( \overline{d}\right) -\Delta \epsilon, 
\label{Lamb detune}
\end{equation}

\noindent where $\overline{d}$ is the mean particle diameter. Using 
these relationships, we assume a distribution in detuning frequency given by
\begin{equation}
P(\nu )\equiv \frac{e^{-\left( \nu -\overline{\nu }\right) ^{2}/\sigma ^{2}}}{\sqrt{\pi }\sigma }.
\end{equation}
\noindent\ The ensemble-averaged population of electronic state $\left|\alpha\right\rangle$
is then given by 

\begin{equation}
\overline{N}_{\alpha }(t)\equiv \int _{-\infty }^{\infty }d\nu \, P(\nu )\, N_{\alpha }(t).
\label{average population}
\end{equation}

The behavior of $\overline{N}_{2}(t)$ at short times is sinusoidal 
with a decaying envelope dependent on the specifics of the particle size distribution.
We will give examples of the short-time dynamics below.

The long-time behavior of $\overline{N}_{2}(t)$ can be obtained analytically
by an asymptotic expansion. We begin by writing (\ref{average population}) as

\begin{equation}
\overline{N}_{2}(t)=\overline{N}_{2}(\infty )+\frac{\Omega _{\rm {res}}^{2}}{2\sqrt{\pi }\sigma }\rm Re\, \it I(t),
\end{equation}

\noindent where

\begin{equation}
\overline{N}_{2}(\infty )\equiv 1-\frac{\Omega _{\rm {res}}^{2}}{2\sqrt{\pi }\sigma }\int _{-\infty }^{\infty }d\nu \, \frac{e^{-\left( \nu -\overline{\nu }\right) ^{2}/\sigma ^{2}}}{\Omega _{\rm {res}}^{2}+\nu ^{2}}
\end{equation}

\noindent is a constant independent of time, and

\begin{equation}
I(t)\equiv \int _{-\infty }^{\infty }d\nu \, \frac{e^{-\left( \nu -\overline{\nu }\right) ^{2}/\sigma ^{2}}}{\Omega _{\rm {res}}^{2}+\nu ^{2}}\, e^{it\sqrt{\Omega _{\rm {res}}^{2}+\nu ^{2}}},
\end{equation}

\noindent where 

\begin{equation}
\Omega_{\rm {res}} \equiv 2g
\label{Res Rabi}
\end{equation}

\noindent is the \emph{resonant} Rabi frequency. It will turn out that the constant 
$\overline{N}_{2}(\infty)$ is simply the value of $\overline{N}_{2}(t)$ in the $t\rightarrow \infty$
limit. Values of this plateau population are given in Table 1.

\begin{table}[!h]
{\centering \begin{tabular}{cc|ccc}
&
&
&
\( \overline{\nu }/\Omega _{\rm res} \)&
\\
&
&
\( 0.0 \)&
\( 1.0 \)&
\( 2.0 \)\\
\hline 
&
\( 0.1 \)&
\( 0.50 \)&
\multicolumn{1}{c}{\( 0.74 \)}&
\( 0.90 \)\\
\cline{3-3} \cline{4-4} \cline{5-5} 
\( \sigma /\Omega _{\rm res} \)&
\( 0.5 \)&
\multicolumn{1}{|c}{\( 0.54 \)}&
\multicolumn{1}{c}{\( 0.73 \)}&
\multicolumn{1}{c}{\( 0.89 \)}\\
\cline{3-3} \cline{4-4} \cline{5-5} 
&
\( 1.0 \)&
\( 0.62 \)&
\( 0.73 \)&
\( 0.87 \)\\
\end{tabular}\par}

\caption{{ Plateau population values for \protect\( \overline{N}_{2}(t)\protect \)
as \protect\( t\rightarrow \infty \protect \). \protect\( \Omega _{\rm {res}}=2g\protect \)
is the resonant Rabi frequency.}\small }
\end{table}

The integral $I(t)$ can be evaluated at large times by analytically continuing
$\nu$ into the complex plane and expanding around the saddle-point at
$\nu=0$. This leads to the asymptotic result

\begin{equation}
I(t)\approx e^{-\overline{\nu }^{2}/\sigma ^{2}}\sqrt{\frac{2\pi }{t\Omega ^{3}_{\rm {res}}}}\, e^{i(t\Omega _{\rm {res}}+\pi /4)},
\end{equation}

\noindent and hence

\begin{equation}
\overline{N}_{2}(t)\approx \overline{N}_{2}(\infty )+\sqrt{\frac{1}{2}}\, \frac{e^{-\overline{\nu }^{2}/\sigma ^{2}}}{\sigma /\Omega_{\rm{res}} }\, \frac{\cos (t\Omega _{\rm {res}}+\pi /4)}{\sqrt{t \Omega_{\rm{res}}}}\qquad (t\rightarrow \infty ).
\label{asymptotic soln}
\end{equation}

Note that at long times the population oscillates at the \emph{resonant} Rabi
frequency, independent of the mean detuning $\overline{\nu }$. This occurs
because the higher-frequency components tend to average out faster. However,
the amplitude of the asymptotic oscillations decreases with $\overline{\nu }$.

To illustrate the short-time behavior of $\overline{N}_2(t)$ as a function of 
$\overline{\nu}$ and $\sigma$, we present several plots showing $\overline{N}_2(t)$
as a function of $t\Omega_{\rm{res}}$.  In choosing values for $\overline{\nu}$ and
$\sigma$, we attempted to cover the entire range of these parameters
which could be reasonably addressed in the RWA.

In Fig.~1, we study the resonant case where
$\overline{\nu}=0$.  By choosing three values of $\sigma$ we are able to directly
observe the effects that it has on the envelope function. As $\sigma$ increases, the
envelope function decays faster and the time at which asymptotic
behavior becomes observable decreases. In this case of $\overline{\nu}=0$, changing
$\sigma$ has little effect on the frequency at which $\overline{N}_2(t)$ oscillates.

\begin{figure}[!h]
{\par\centering \includegraphics{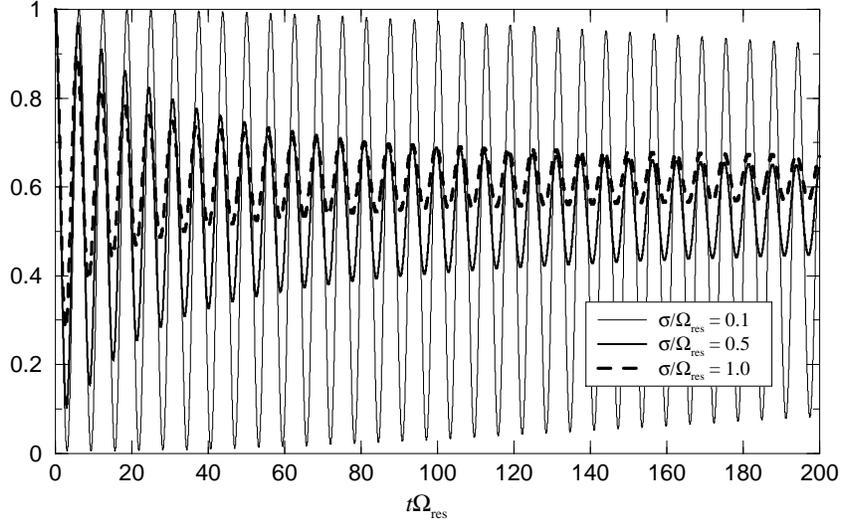} \par}
\caption{Electronic population in state $\left| 2\right\rangle$ for
$\overline{\nu}=0.0$.}
\end{figure}

Fig.~2, where $\overline{\nu}=\Omega_{\rm{res}}$, illustrates 
the effects of an intermediate value of $\overline{\nu}$.  The 
short-time envelope function decays faster and the plateau population is increased.
As in Fig.~1, increasing $\sigma$ increases the rate 
at which the short-time envelope function decreases, but here it can be seen to
lower the frequency of oscillation in $\overline{N}_2(t)$. Unlike Fig.~1, Fig.~2
clearly shows how increasing $\sigma$ up to $\Omega_{\rm{res}}$ increases the 
amplitude of oscillations at large times [see Eqn. (\ref{asymptotic soln})].   

\begin{figure}[!h]
{\par\centering \includegraphics{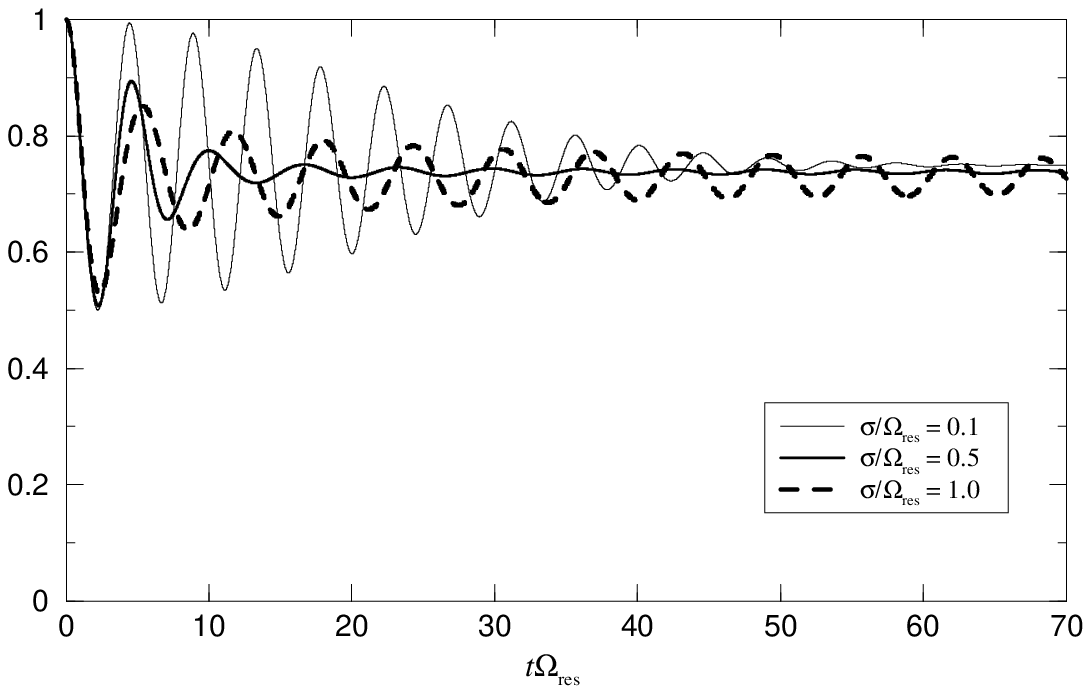} \par}
\caption{Electronic population in state $\left| 2\right\rangle$ for
$\overline{\nu}=\Omega_{\rm{res}}$.}
\end{figure}

The last plot, Fig.~3, illustrates the limits of the RWA. The effects of increasing
$\overline{\nu}$ are now taken to the extreme: the short-time envelope functions
decay very rapidly, making asymptotic behavior apparent at early times, and the
plateau population approaches unity.  Altering $\sigma$ has similar effects as
in Figs. 1 and 2, only more pronounced. [$\overline{N}_2(t)$ for 
$\sigma/\Omega_{\rm{res}}=0.1$ levels off to its plateau population shortly past 
$t\Omega_{\rm{res}}=50$.]

\begin{figure}[!h]
{\par\centering \includegraphics{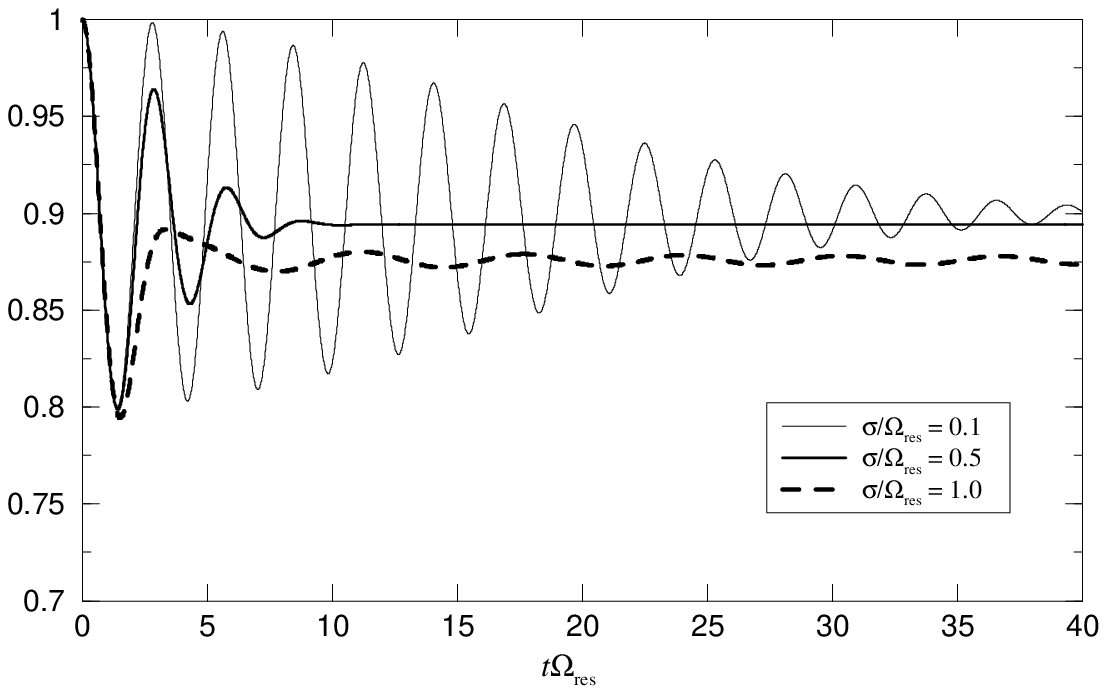} \par}
\caption{Electronic population in  state $\left| 2\right\rangle$ for
$\overline{\nu}/\Omega_{\rm{res}}=2.0$.}
\end{figure}

\section{Population dynamics in Y$_2$O$_3$:Eu$^{3+}$ nanoparticles}

Here we use the results of Section 2 to address possible future experiments.
We examine the case of Y$_2$O$_3$:Eu$^{3+}$ nanoparticles, specifically the electron
dynamics between two crystal-field split $^5$D$_1$ levels separated by 
$\Delta\epsilon=3$~cm$^{-1}$, such as in the experiment by Yang \emph{et al.}~\cite{Yang etal}. 
To make quantitative predictions for such nanoparticles, the only parameters that need to 
be specified are the mean detuning frequency $\overline{\nu}$, and the standard deviation 
in detuning $\sigma$. We also need the electron-phonon interaction strength $g$. Our 
treatment requires that $\overline{\nu}$ be not much greater than $\Omega_{\rm{res}}$.

In the case of  Y$_2$O$_3$:Eu$^{3+}$ nanoparticles, we can estimate the electon-phonon 
interaction strength by using the experimentally observed bulk phonon emission rate 
\cite{Yang etal}. We obtain a value for $g_{\rm{micron}}$, the electron-phonon interaction 
strength in micron-size particles, and then scale this value to get $g$ for the nanoparticles. 
The scaling is achieved by assuming that $g$ varies with energy and system volume as it does
in a bulk crystal. In principle, this method is only correct between two systems with 
continuous vibrational spectra. However, using this technique should provide a reasonable
estimate of $g$ in the nanoparticle.

In the micron-scale crystal, Fermi's golden rule states that for an electron-phonon
interaction of the form given in Eqn.~(\ref{hamiltonian}), the phonon emission rate is

\begin{equation}
\tau ^{-1}=\frac{2\pi }{h}g^{2}\, \Gamma (\Delta\epsilon )\, V,
\end{equation}

\noindent where \( V \) is the crystal volume and $\Delta\epsilon$ is the electronic
level spacing. \( \Gamma (\epsilon ) \), the phonon DOS (per volume) as a function of 
phonon energy \( \epsilon  \), is defined by

\begin{equation}
\Gamma(\epsilon) \equiv \frac{1}{V}\sum_{n}\delta(\epsilon - \hbar\omega_{n}),
\end{equation}

\noindent where $\omega_n$ are the phonon frequencies, and in a bulk crystal at low energies

\begin{equation}
\Gamma(\epsilon) = \frac{\epsilon^2}{2\pi^2\hbar^3\overline{v}^3},
\end{equation}

\noindent where

\begin{equation}
\overline{v} \equiv \left( \frac{1}{3} \sum _{\lambda }\frac{1}{v_{\lambda }^{3}}\right) ^{-1/3}
\end{equation}

\noindent is a branch-averaged sound velocity. 

The lifetime in the micron-sized crystal was observed to be $\tau=220$~ns \cite{Yang etal}, and
in  Y$_2$O$_3$ the sound velocities are approximately \cite{Proust},

\begin{equation}
v_{\rm{l}} = 6.7 \times 10^5\: \rm{ cm\: s^{-1}},
\end{equation}
\begin{equation}
v_{\rm{t}} = 4.3 \times 10^5\: \rm{ cm\: s^{-1}},
\end{equation}

\noindent and

\begin{equation}
\overline{v} = 4.7 \times 10^5\: \rm{ cm\: s^{-1}}.
\end{equation}

\noindent $g$ is then given by

\begin{equation}
g=\frac{\pi\hbar^{2}\overline{v}^{3/2}}{\Delta\epsilon}\sqrt{ 2 \tau ^{-1} \over 3 V}.
\end{equation}

The mean diameter of these nanoparticles was about 5 $\mu$m, leading to
\( g_{\rm{micron}}=2.04\times 10^{-6} \)~cm\( ^{-1} \).

In a bulk crystal (with deformation-potential coupling), $g$ varies as $(\epsilon/V)^{1/2}$, 
so we scale $g_{\rm{micron}}$ in $\epsilon$ and $V$ to get a value for $g_{\rm{nano}}$, 
the electron-phonon interaction strength in the nanoparticle,

\begin{eqnarray}
g_{\rm{nano}} & = & \left( \frac{10\: \rm cm^{-1}}{3\: \rm cm^{-1}}\right)^{1/2} \left( \frac{13\: \rm {nm}}{5\: \mu \rm {m}}\right) ^{-3/2}g_{\rm{micron}},\\
 & = & 2.8\times 10^{-2}\: \rm {cm^{-1}}.
\end{eqnarray}

\noindent Then the resonant Rabi frequency, as defined in Eqn.~(\ref{Res Rabi}), is about

\begin{equation}
\Omega_{\rm{res}}=5.6 \times 10^{-2}\: \rm{cm^{-1}}.
\end{equation}

In the experiment of Yang \emph{et al.}, the mean particle diameter was 13~nm and the
standard deviation in particle size was 5~nm \cite{Yang etal}.  By Eqns.~(\ref{Lamb mode}) and
(\ref{Lamb detune}), these values yield $\overline{\nu}=71\Omega_{\rm{res}}$ and 
$\sigma=44\Omega_{\rm{res}}$. This detuning is large enough that higher-energy vibrational 
modes can no longer be ignored, making our single mode treatment inapplicable. Furthermore, 
these values are too far off resonance to be reliably addressed by the RWA.

To observe the behavior predicted in Section 2 for Y$_2$O$_3$:Eu$^{3+}$ nanoparticles, the
mean diameter should be about 46~nm, and standard deviation in diameter no more than about 
3~nm [see Eqns.~(\ref{Lamb mode}) and (\ref{Lamb detune})].

\section{Discussion}

We have shown that the unavoidable size-dispersion in a collection of nanoparticles 
can effect the ensemble-averaged electronic population dynamics considerably, affecting
even the oscillation frequency at long times. However, our analysis has been restricted
to near-resonant conditions in an idealized single-mode nanocrystal. A comprehensive 
theory of electron-phonon dynamics in nearly isolated nanoparticles will have to
address a number of additional issues.

First, the vibrational modes of a nanoparticle will be broadened due to intrinsic 
mechanisms and from interaction to the environment. For example, anharmonicity will 
broaden the Lamb mode at finite temperature \cite{Markel and Geller}, as will mechanical
coupling to a substrate or to a cluster of other nanoparticles \cite{Patton and Geller}. 
This broadening will cause the envelope of the population oscillations to decay exponentially 
at long times. For weak damping, however, we do not expect any qualitative changes in the
short-time dynamics.

Second, nanoparticles will generally have large detuning frequencies, making the inclusion 
of higher-frequency modes necessary, and making the RWA invalid. One way to include these 
effects would be to do exact-diagonalization studies in models with a truncated Hilbert space;
for example, having bounded phonon occupation numbers for one or more modes. 

Finally, it will be necessary to understand the effects of vibrational-mode degeneracy,
which has been neglected here. The presence of degeneracy means that the Lamb mode has
a vector-like internal degree of freedom, the components of the vector describing the
phonon amplitude in each branch. Although there has been work done in the quantum optics
literature on the multi-mode generalizations of the our Hamiltonian (\ref{hamiltonian}),
that work has focused on the two-photon resonance case where $\Delta \epsilon = 2 \omega_0$
\cite{Caves and Schumaker, Gou}. One way to approach the degenerate case nonperturbatively 
would be to assume $N$ degenerate modes in the limit of large $N$. By analogy with other 
quantum systems with internal degrees of freedom, like particles with $N$ spin components, 
it is reasonable to expect that quantum effects will be diminished in the large $N$ limit. 
In a nanoparticle this would suggest a suppression of the Rabi oscillations.

\section{Acknowledgements}

This work was supported by a Research Innovation Award from the Research Corporation.
It is a pleasure to thank Bill Dennis, Richard Meltzer, and Kelly Patton for useful
discussions.



\begin{thebibliography}{1}
\bibitem{vacuum footnote} Vacuum refers to the case where no phonons, thermal or
otherwise, are present in the initial electron-phonon state.
\end{thebibliography}
\end{document}